\documentclass[num-refs]{wiley-article}

\usepackage{siunitx}
\usepackage{xr}
\usepackage{tikz}
 \usepackage{tkz-graph}
  \usepackage{pgf}
\usepackage{amsmath}
\usepackage{float}
\usepackage{bbm}
\usepackage{enumerate}
\usepackage[english]{babel}
\usepackage{bbm}
\usepackage{blindtext}
\usepackage{dsfont}
\usepackage{algorithm}
\usepackage[noend]{algpseudocode}
\usepackage{lipsum}
\usepackage{amssymb}
\usepackage[inline]{enumitem}
\usepackage{url} 
\usepackage{multirow}
\usepackage[table]{xcolor}
\usepackage{accents}
\usepackage{longtable}

\makeatletter
\newcommand*{\addFileDependency}[1]{
  \typeout{(#1)}
  \@addtofilelist{#1}
  \IfFileExists{#1}{}{\typeout{No file #1.}}
}
\makeatother

\newcommand{\be}{\begin{eqnarray}}
\newcommand{\ee}{\end{eqnarray}}
\newcommand{\var}{\mbox{Var}}

\usepackage{scalerel,stackengine}
\stackMath
\newcommand\reallywidehat[1]{%
\savestack{\tmpbox}{\stretchto{%
  \scaleto{%
    \scalerel*[\widthof{\ensuremath{#1}}]{\kern-.6pt\bigwedge\kern-.6pt}%
    {\rule[-\textheight/2]{1ex}{\textheight}}
  }{\textheight}%
}{0.5ex}}%
\stackon[1pt]{#1}{\tmpbox}%
}
\papertype{Original Article}
\paperfield{Journal Section}

\title{Explained Variation under the Additive Hazards Model}

\author[1]{Denise Rava}
\author[1,2]{Ronghui Xu}

\affil[1]{Department of Mathematics, University of California, San Diego}
\affil[2]{Department of Family Medicine and Public Health, University of California, San Diego}

\corraddress{Ronghui Xu, 9500 Gilman Drive, Mail code 0112, La Jolla, CA 92093.}
\corremail{rxu@ucsd.edu}

\begin{document}

\maketitle

\begin{abstract}
We study explained variation under the additive hazards regression model for right-censored data. 
We consider different approaches for developing such a measure, and focus on one that  estimates the proportion of  variation in the failure time explained by the covariates. We study the properties of the measure both analytically, and through extensive simulations. 
We apply the measure to a well-known survival data set as well as the linked Surveillance, Epidemiology and End Results (SEER)-Medicare database for prediction of mortality in early-stage prostate cancer patients using high dimensional claims codes.

\end{abstract}
\newtheorem{thm}{Theorem}

\section{Introduction}

The additive hazards model   \cite{Aalen, aalen1989linear} has received increasing attention lately
for the analysis of censored survival data. 
 It is not just an alternative to the more widely used Cox model when the  proportional hazards assumption is violated; 
it has also been argued to be more suitable for causal inferences in estimating treatment effects because the Cox model 
 is not collapsible \cite{mart:vans}. 
In contrast, the additive hazards model behaves mostly like a linear model including  collapsibility, 
 in the sense that one can integrate out an independent covariate from the model and still end up with an additive hazards model, with the same regression coefficients for all the other covariates. 
For this reason it has been used in the development of instrumental variable approaches for  survival data including competing risks \cite{tchetgen2015instrumental, li2015instrumental, zheng2017instrumental, jiang2018two, brueckner2019instrumental,andrew}. 
The collapsibility as well as other behaviors similar to a linear model, has also enabled the additive hazards model to be used in mediation analysis of survival data \cite{fosen2006dynamic,martinussen2010dynamic, martinussen2011estimation,vanderweele2013unmeasured,aalen2020time}. 
In addition, doubly robust methods have been developed for estimating treatment effects and applied in practice under the  additive hazards model including for optimal treatment regimes \cite{wang2017doubly,kang2018estimation, blomberg2019long}, 
while the noncollapsibility of the Cox model presents an obstacle in the development of doubly robust method when confounders are present \citep{dukes:19}.

Estimation and inference procedures have been well developed and implemented under the additive hazards model (eg. R package `timereg'), and diagnostic methods have also been proposed  \cite{diagn6, diagn7, timereg}. 
However, another important aspect  as the model becomes more widely used, is explained variation or measures of predictability, often referred to as $R^2$. 
\citet{oq:xu:12} provide detailed illustrations of how 
such measures are used to evaluate the clinical importance of prognostic factors.
   \citet{muller2008quantifying}  and  \citet{hielscher2010prognostic} explored the use of $R^2$ measures in genetic studies to quantify the impact of genetic variants or high dimensional  gene expression on survival phenotypes, while \citet{pres:molen:2011} applied them to surrogate evaluation. 
  Very recently applications of measure of dependence to ultrahigh dimensional variable screening were explored in \citet{kong2019composite}.
In the context where the estimation of treatment effect is of primary concern, following the fit of the additive hazards models it is also natural to provide estimates of predicted survival given the covariates \cite{andrew}. However, measures of explained variation have not been examined under the additive hazards model to our best knowledge.

Explained variation has been well studied 
in the literature under the Cox regression model for right-censored data.
Kent and O'Quigley \citep{KO} first defined a measure of dependence for censored survival data, making use of the Kullback-Leibler information gain. It is based on the conditional distribution of the time to event random variable $T$ given the covariates $Z$. 
A later work by Xu and O'Quigley  \citep{ZgivenT} considered instead the conditional distribution of  $Z$ given  $T$, 
using also the information gain. This latter measure can be readily extended to time-dependent covariates. 
A simple approximation to this second measure  was described in \citet{CoxRho2}, which can be easily computed using the partial likelihood ratio test statistic following the fit of the Cox model. \citet{pres:molen:2011} advocated for these  information gain based measures. 

Another approach to defining explained variation 
 makes use of the residuals. This  originated from  the  $R^2$ under the  linear regression model, which can be written as one minus the ratio of the residual sum of squares over the total sum of squares. It is also well-known that these two sums of squares estimate the residual variance and the total variance, respectively. 
O'Quigley and Flandre \citep{PNAS} proposed to  use the Schoenfeld residuals under the Cox model, in a similar way to the $R^2$ under linear regression. 
It has been shown that when the Cox model appears to be a reasonably fit to the data, this measure and the one above based on information gain, tend to give comparable quantifications of explained variation \cite{oq:xu:12}. 
 
 Other approaches have also been considered in the literature 
 for right-censored data. 
\citet{Imp} proposed to compute the correlation coefficients between the failure rankings and the covariates,  using multiple imputation to handle the censored data. We note  that inference under the Cox model is only based on the ranks of the failure times, hence nonparametric correlation coefficients like Kendall's tau or Spearman correlation might be considered. However, as it is known and we also elaborate below, inference under the additive hazards  model is not rank based. 

Finally and not restricted to the survival context, previous experiences in describing explained variation outside the classic linear model have also considered the direct decomposition of the total variance in the outcome, and quantifying the proportion  that is explained by the covariates. Depending on the model, this can sometimes be a straightforward approach, such as under the linear mixed effects model \cite{LmmR2, honer:xu:16}, or under the accelerated failure time (AFT) models \cite{chan:etal:2016}. 

In this work we consider the semiparametric additive hazards  model. 
We aim to quantify the explained variation under this model. It turns out that the last approach described above, i.e.~the direct decomposition of the total variation into components of explained and unexplained (or residual) variation, is easily computable as well as interpretable  under the additive hazards model. In the following we will first focus on its development, investigate its properties, and illustration how it might be used in practice 
to quantify the predictive power of a set of prognostic variables, and also for use in  variable selection procedures. 
We will defer discussion to the end of the paper  why some of the other approaches described above do not work under  the additive hazards  model.

The rest of the paper is organized as follows. 
After a review of the semiparametric additive hazards model and its inference  in the next section, we describe explained variation and its estimator in section \ref{estimator}. In section \ref{effect}, we study the properties of the measure, both the population and the sample-based versions. Section \ref{simulations} further explores the behavior of the measures using simulation,  under different censoring scenarios, different covariate distributions, different  baseline hazard functions, and beyond. 
We apply the measure to real data sets in Section \ref{seer-sec},  and we conclude with discussion in the last section.

\section{Semiparametric Additive Hazards Model }\label{model}

Let $T$ be the failure time random variable of interest, $Z$ be a vector of covariates, and $C$ be the censoring time random variable. Let $X=\min{\left(T,C\right)}$  and $\delta=
I{(T\leq C)}$ where $I(\cdot)$ is the indicator function. We observe a random sample $(X_i,Z_i,\delta_i)$, $i=1,\ldots,n$. 
The semiparametric additive hazards model  \cite{ly} 
assumes that the conditional hazard function 
\be\label{add_const}
\lambda(t|Z) = \lambda_0(t) + \beta^\top Z,
\ee
where $\lambda_0(t)$ 
 is the baseline hazard and $\beta$ is a vector of regression effects. 
We will also use the counting process notation: 
 $N(t)= I{\left\{X\leq t, \delta=1\right\}}$ and $Y(t)= I{\left\{X\geq t\right\}}$ are the counting process of events and the at-risk process,  respectively. 
 
 Under  model \eqref{add_const}, an estimator for $\beta$ was proposed by \citet{ly}: 
\be\label{beta}
\hat{\beta}=\left[\sum_{i=1}^n\int_0^\infty Y_i(t)\left\{Z_i-\bar{Z}(t)\right\}^{\otimes2}dt\right]^{-1}\left[\sum_{i=1}^n\int_0^\infty \left\{Z_i-\bar{Z}(t)\right\} dN_i(t) \right],
\ee
where 
$ 
\bar{Z}(t)=\sum_{i=1}^n{Y_i(t)Z_i} / {\sum_{i=1}^nY_i(t)}
$. 
We note that unlike under the Cox model, the above estimator of  $\beta$ is not rank based in that it depends on the values of $X_i$'s beyond their ranks in the data set. 
It can be shown that, if $g(\cdot)$ is a strictly increasing function, then $g(T)$ in general no longer follows a semiparametric additive hazards model. 
In the special case 
where $g$ is multiplication by a  constant $c>0$, then
$\tilde{T}=cT$ still follows a semiparametric additive hazards model, but the regression coefficient is rescaled by c: $\tilde\beta = \beta/c$. 

The cumulative baseline hazard function $\Lambda_0(t)=  \int_0^t \lambda_0(u) du$ is estimated by 
\be\label{cumbase}
\tilde{\Lambda}_0(t)=\int_0^t\frac{\sum_{i=1}^n\left(dN_i(u)-Y_i(u)\hat{\beta}(u)^\top Z_idu\right)}{\sum_{j=1}^nY_j(u)}.\ee
In the following we write out  the integral in \eqref{cumbase}, which is not a step function.
Denote the $K$ ordered distinct observed failure times $t_1< ... < t_K $. We have
for $k=1, ..., K$:
\be
\tilde{\Lambda}_0(t_k) = \sum_{l=1}^{k} \frac{ \delta_{l}d_{l}}{r_{l} } -
\sum_{l=1}^k \hat{\beta}^\top\bar{Z} \left(t_{l}\right)\left(t_{l}-t_{l-1}\right), 
\ee
where $ d_{l}$ and $r_{l}$ are the number of events and number at risk at time $t_l$, respectively.
In addition, for any $t_k \leq t< t_{k+1}$,
\be
\tilde{\Lambda}_0(t)
 = \sum_{l=1}^{k} \frac{ \delta_{l}d_{l}}{r_{l} } -
\sum_{l=1}^k \hat{\beta}^\top\bar{Z} \left(t_{l}\right)\left(t_{l}-t_{l-1}\right)
-\hat{\beta}^\top \bar{Z} \left(t_{k+1}\right)\left(t- t_{k}\right).
\ee
The resulting estimated survival function $\tilde{S}(t | z)=\exp{(-\tilde{\Lambda}_0(t)-\hat{\beta}\top zt)}$ is not guaranteed to be non-increasing; therefore we make use of the following adjusted version \cite{ly}: $
\hat{S}(t | z)=\min_{s\leq t}\left\{\tilde{S}(s |z)\right\}
$. 
The adjusted version $\hat{S}$ is asymptotically equivalent to $\tilde S$, and the process $\sqrt{n}(\hat{S}(\cdot | z)-{S}(\cdot | z))$ converges wealy to a zero-mean Gaussian process \cite{ly}.
We note that taking minimum over $s\leq t $ leads to no closed-form expression and the quantity needs to be computed numerically. However, it is imperative that we work with a proper distribution or equivalently, survival, function, in order to estimate the moments below.

\section{Explained Variation}\label{estimator}

The explained variation, as described in the survival context by \citet{oq:xu:12}, 
can be defined as
\be\label{omega}
\Omega^2 = 1-\frac{ E \left\{\var(T\mid Z )\right\}}{\var(T)} = \frac{ \var\left\{E(T\mid Z )\right\}}{\var(T)}.
\ee
This is consistent with the regression setting of model \eqref{add_const} for the conditional distribution of $T$ given $Z$, as the proportion of variation of $T$ explained by $Z$ out of the total variation of $T$. 
As pointed out in \citet{o2008proportional} page 33, by virtue of the Chebyshev-Bienayme inequality, the variance can be seen as a measure of predictability, and therefore 
the explained variation may also have an interpretation as predictability.

In practice for survival studies, there is often a finite upper bound of time $\tau $ due to administrative censoring, so that all the observable data are conditional upon $  T< \tau$.  We then define
\be\label{omega_tau}
    \Omega^2_\tau=1-\frac{ {E}\left\{\mbox{Var}(T\mid Z, T< \tau)\right\} }{ \mbox{Var}(T \mid T< \tau) }
    = \frac{ \mbox{Var}\left\{ {E}(T\mid Z, T< \tau)\right\} }{ \mbox{Var}(T \mid T< \tau) }.
\ee
Obviously when there is no censoring, $\Omega^2 = \Omega^2_\infty$; and 
in the following for uniformity of notation,  we allow $\tau \leq \infty$.

We can estimate directly the quantities in \eqref{omega_tau} under model \eqref{add_const}.
To estimate $E\left\{\mbox{Var}(T\mid Z, T< \tau )\right\}$ or \\
$\mbox{Var}\left\{E(T\mid Z, T< \tau)\right\} $, we first integrate with respect to an estimated distribution of $T$ given $Z$ and $ T< \tau $:
\be
\hat{S}(t\mid Z, T< \tau) =\frac{\hat{S}(t\mid Z)-\hat{S}(t_K\mid Z)}{1-\hat{S}(t_K\mid Z)} \mathbb{1}\left\{t\leq t_K\right\}
\ee
We then integrate with respect to $\mathbb{P}_n$,  the empirical distribution of $Z$. 
Denote the resulting estimates 
$ {E}_n \left\{\widehat{\mbox{Var}} \left(T \mid Z, T< \tau\right)\right\} $ and 
$  \mbox{Var}_n \left\{\hat{{E}} (T\mid Z, T< \tau)\right\} $, respectively. 
For example, 
\be\label{varTZ}
  {E}_n \left\{\widehat{\mbox{Var}} \left(T \mid Z, T< \tau\right)\right\}
    =\frac{1}{n}\sum_{i=1}^n\left[ {\hat{E}\left(T^2\mid Z_i, T< \tau\right)}-\left\{ {\hat{E}\left(T\mid Z_i, T< \tau\right)}\right\}^2\right],
\ee
where the expressions for the quantities in the right-hand side above are given later in the section. 

To estimate $\mbox{Var}(T \mid T\leq \tau)$,  we can use 
\be\label{varTvar}
    \widehat{\mbox{Var}} \left(T \mid T< \tau\right) = {\hat{E}\left(T^2\mid  T< \tau\right)}-\left\{ {\hat{E}\left(T\mid T< \tau\right)}\right\}^2.
\ee
In order  to estimate the marginal survival function, we may use the nonparametric Kaplan-Meier (KM) estimator. Alternatively, 
 we may use:
\be\label{marg}
\hat{S}(t \mid T< \tau)=\frac{1}{n}\sum_{i=1}^n\hat{S}(t\mid Z_i, T< \tau).
\ee
It can be shown that, if  \eqref{marg} is used in estimating the expectations in \eqref{varTvar},
then we have the following decomposition:
\be\label{SSdecomp}
\widehat{ \mbox{Var}} (T\mid T< \tau)  =
{ {E}_n \left\{\widehat{\mbox{Var}} \left(T \mid Z, T< \tau\right)\right\} }
+ { \mbox{Var}_n \left\{\hat{E}(T\mid Z, T< \tau)\right\} }. 
\ee

Combining  all of the above, we obtain $ R^2_\tau $ as a consistent estimator of  $\Omega^2_\tau$ under model \eqref{add_const}:
\be\label{r2}
    R^2_\tau=1-\frac{ { {E}_n\left\{\widehat{\mbox{Var}} \left(T\mid Z, T< \tau\right)\right\}}}{\widehat{\mbox{Var}} \left(T \mid T< \tau\right)} 
    = \frac{ {\mbox{Var}_n \left\{\hat{E}\left(T\mid Z, T< \tau\right)\right\}}}{\widehat{\mbox{Var}} \left(T \mid T< \tau\right)} .
\ee
We also denote $R^2 = R^2_\infty $ when $\tau=\infty$. 

Finally, to compute the quantities in \eqref{r2}, 
we have: 
\be
{\hat{E}\left(T \mid z, T< \tau\right)}&=&
\int_{0}^{\tau} \hat{S}\left(t\mid z, T< \tau\right)dt \nonumber\\
&=&\frac{1}{1-\hat{S}\left(t_K\mid z\right)}
 \int_{0}^{t_K} \hat{S}\left(t\mid z\right)dt-\frac{1}{1-\hat{S}\left(t_K\mid z\right)} \hat{S}\left(t_K\mid z\right) t_K,
\ee
and
\be
{\hat{E}\left(T^2 \mid z, T< \tau\right)}&=&
2 \int_{0}^{\tau} t \cdot \hat{S}(t\mid z, T< \tau) dt
\nonumber\\
&=&\frac{2}{1-\hat{S}\left( t_K \mid z\right)}
\int_{ 0}^{ t_K} t\hat{S}\left( t \mid z\right)dt
-\frac{1}{1-\hat{S}\left( t_K \mid z\right)}\hat{S}\left( t_K \mid z\right) t_K^2.
\ee
Since there is no closed-form expression for $\hat{S}(t \mid Z)$, the integrals in the  above are computed using the trapezoidal rule.  We  partition the interval $[0,\tau]$ first using $t_1,\ldots,t_K $; additional points are added to create a  grid no wider than 0.01 between  two adjacent points. We then use an iterative 
halving process, i.e.~adding the midpoints between any two adjacent points to the grid,  until the change in the resulting $R^2_{\tau}$  is less than  0.01 in absolute value.

The quantities in 
$\widehat{ \mbox{Var}} (T\mid T< \tau) $ can be computed in a similar fashion  using  \eqref{marg}.

\section{Properties of $\Omega^2$ and $R^2$}\label{effect}

The desirable properties of a measure of explained variation are best understood under a linear regression model, including: 
1) it lies between zero and one;
2) it takes the value zero when there is no regression effect;
3) it increases with the strength of the regression effect;
4) it tends to one as the regression effect tends to infinity;
5) it is invariant under certain transformations of the dependent and independent variables, depending on the model.
For the last property, the transformation is linear under the linear regression model, and is rank-preserving for the failure time under the semiparametric Cox regression model \cite{oq:xu:12}.

In the following we investigate if the above properties hold for the measures defined in the last section.

\begin{itemize}

\item The facts that $0\leq \Omega^2_\tau \leq 1$ and $0\leq R^2_\tau \leq 1$ follow immediately from their definitions \eqref{omega_tau} and \eqref{r2}, assuming that the latter is estimated using \eqref{marg}.

\item When $\beta=0$, $\Omega^2_\tau =0$ because
 independence between $T$ and $Z$ implies that $\mbox{Var}\{E(T \mid T<\tau,Z)\}=\mbox{Var}\{E(T \mid T<\tau)\}$.
 Also $R^2_\tau =0$ if it happens that the estimated coefficient $\hat{\beta}=0$. Otherwise, the sample based measure $R^2_\tau>0$, but is expected to be small since it is a consistent estimate of $\Omega^2_\tau =0$. 

\item It is analytically difficulty to prove that $\Omega^2_\tau$ increases with $|\beta|$ in general. However, for simpler settings such as a binary $Z$ and $\tau=\infty$, we can prove it analytically and this is given in the Appendix. For more general settings, we illustrate this via simulation. 

\item It has been known that the quantity $\Omega^2$ defined in \eqref{omega} can be bounded strictly less than one \cite{oq:xu:12}.
For a binary $Z$, if we assume that $T \mid Z=0$ has finite second moment, then we can show by the dominated convergence theorem that:
\be \label{limitform}
\lim_{\beta \rightarrow \infty} \Omega^2_\infty =1 - \frac{\frac{1}{2}\left[2\int_0^\infty t \exp{\left\{ -\Lambda_0(t)\right\}}dt - \left[\int_0^\infty \exp{\left\{ -\Lambda_0(t)\right\}}dt \right]^2\right]}{\int_0^\infty t \exp{\left\{ -\Lambda_0(t)\right\}}dt-\frac{1}{4}\left[ \int_0^\infty \exp{\left\{ -\Lambda_0(t) \right\}}dt\right]^2}.
\ee
For example, when 
$\lambda_0(t)=1$,  $\lim_{\beta \rightarrow \infty} \Omega^2_\infty =0.333$; and this is the exponential case discussed in \citet{oq:xu:12}. 
When $\lambda_0(t)=t$,  $\lim_{\beta \rightarrow \infty} \Omega^2_\infty =0.647$; and when $\lambda_0(t)=1/(2\sqrt{t})$, $\lim_{\beta \rightarrow \infty} \Omega^2_\infty =0.091$. 
Similar calculation can be done for covariates with continuous distribution:
\be \label{limitform}
\lim_{\beta \rightarrow \infty} \Omega^2_\infty =1 - \lim_{\beta \rightarrow \infty}\frac{\int_{\mathcal{Z}}\left[\int_0^\infty2t\exp{\left\{ -\Lambda_0(t)-\beta^\top Zt\right\}}dt-\left[\int_0^\infty\exp{\left\{ -\Lambda_0(t)-\beta^\top Zt\right\}}dt\right]^2\right]g(z)dz}{\int_{\mathcal{Z}}\int_0^\infty2t\exp{\left\{ -\Lambda_0(t)-\beta^\top Zt\right\}}dtg(z)dz-\left[\int_{\mathcal{Z}}\int_0^\infty\exp{\left\{ -\Lambda_0(t)-\beta^\top Zt\right\}}dtg(z)dz\right]^2},
\ee
where $g(Z)$ is the density of the covariates and $\mathcal{Z}$ is their 
 sample space.
This limit may not be equal to one and it depends on the form of $\lambda_0(t)$ and the distribution of  $Z$; for example, when $Z \sim U\left[0,\sqrt{3}\right]$ and 
$\lambda_0(t)=1$,  $\lim_{\beta \rightarrow \infty} \Omega^2_\infty =0.500$.

\item By their definitions 
and simple algebra, it can be shown that $\Omega^2_\tau$ and $R^2_\tau$ are invariant under linear transformations of $Z$ and when $T$ is rescaled by a positive constant. 

\end{itemize}

In summary, we have the following properties: 
\begin{itemize}
\item[1)]  $0\leq \Omega^2_\tau \leq 1$, and $0\leq R^2_\tau \leq 1$; 
\item[2)]  $\Omega^2_\tau =0$ when $\beta=0$, and $R^2_\tau =0$ if  $\hat{\beta}=0$; 
\item[3)]  $\Omega^2_\tau$ increases with $|\beta|$;
\item[4)]  $\Omega^2_\tau$ and $R^2_\tau$ are invariant under any linear transformation of $Z$ and rescaling of $T$. 
\end{itemize}

\section{Simulations}\label{simulations}

In the following we further study the  properties of the measures through simulations.
In addition to the properties mentioned above, we also investigate: 1) the effect of baseline hazard on explained variation; 2) explained variation under nested models. As we have more experience with explained variation under the Cox proportional hazards regression model, we also investigate 3) how the measure compares with a similar one under the Cox model, when both models are valid; and 4) explained variation of $Z$ give $T$, which has been advocated for use under the Cox model.

All simulations below were carried out with sample size 1000, and 100 simulation runs each. All the results are reported as mean with standard deviation (SD) over the simulation runs in $(\cdot)$. As the simulation has been extensive, we have chosen to display the representative scenarios that carry meaningful messages, as opposed to every combination of all possible parameters and settings. 

\subsection{Basic properties }

\subsubsection*{As $|\beta|$ increases }

We first simulated with $\lambda_0(t) =1$ and different values $\beta = $1, 3, 15 and 50, $Z$ from Uniform $[0,\sqrt{3}]$ as well as binary 0,1 with equal probabilities. Note that these two covariate distributions have the same variance 0.25, rendering the measures comparable for a given $\beta$ value. 
The censoring distribution was uniform $[0,\tau]$.
We computed the  $\Omega^2_{\tau}$ values as follows. When there was no censoring we computed it analytically by definition using the fact that $T \sim$ Exponential ($1+\beta Z$). 
When there was censoring, we took  a single large sample size of 100,000, and  used the $R^2_\tau$ value computed with the true $\beta$ and the true $\lambda_0$  to approximate $\Omega^2_{\tau}$.

\begin{figure}[H]
\caption{$\Omega^2_\tau$ and $R^2_\tau$  values for different $\beta$ and $\tau$
 under the model $\lambda(t)=1+\beta Z$. 
}\label{complete}
    \includegraphics[width=1.0\textwidth]{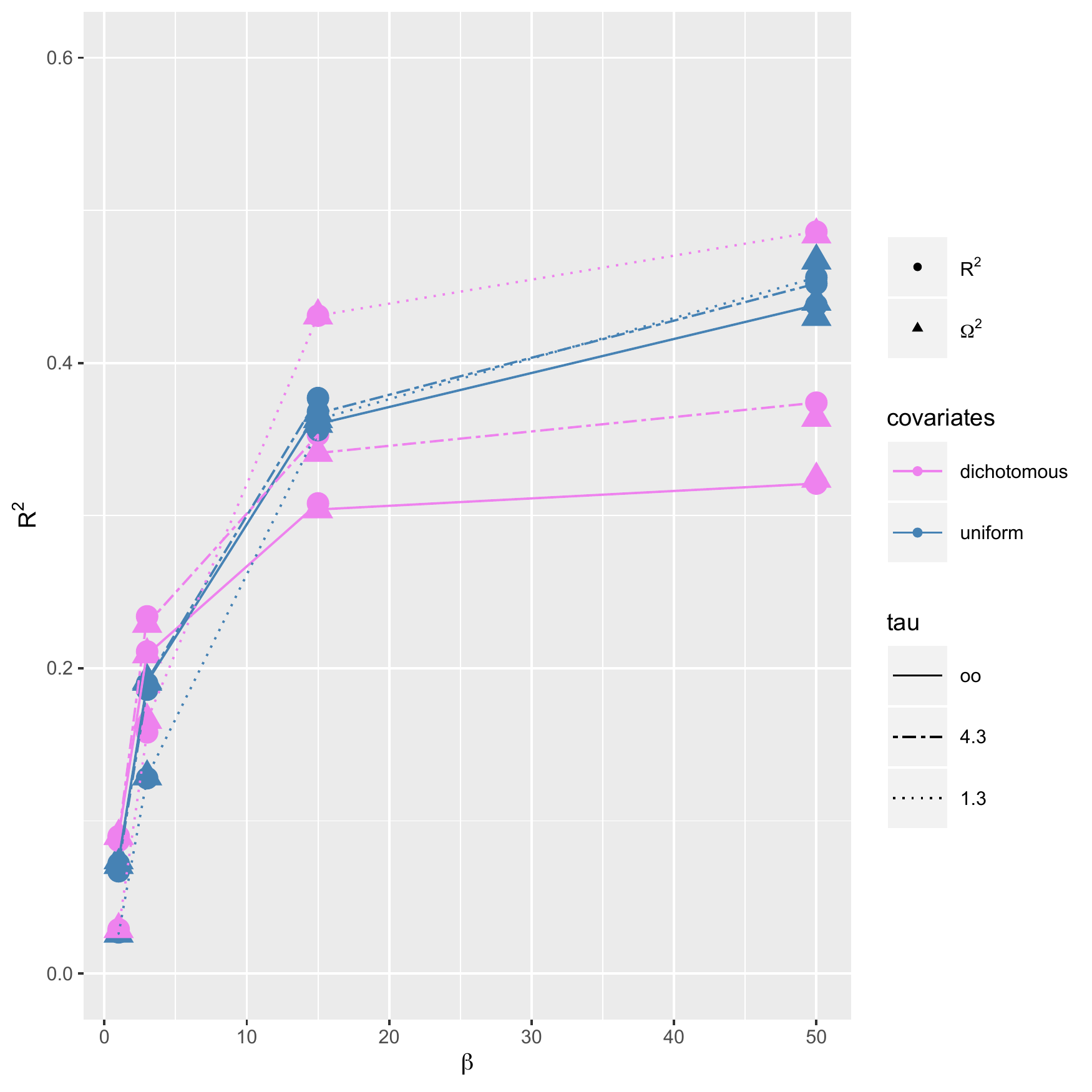}
\end{figure}

From Figure \ref{complete} and Table \ref{fix} we see that $R^2_{\tau}$ and $\Omega^2_{\tau}$ values are close in all cases, both increasing with $|\beta|$ as expected. 
The effect of $\tau$ reflects different follow-up periods, which also leads to different amounts of censoring. 
It is seen that the patterns of change with $\tau$ is different  depending on the  distribution of $Z$.  
It is more pronounced with binary $Z$  especially for that larger $\beta$ values, likely because the censor percentages are much higher  in that case.

\subsubsection*{Effect of $\lambda_0(\cdot)$}

\begin{figure}[H]
\caption{Density of $T$ for each of $Z=0$, 1 groups,
 superimposed with the average $R^2$ values over simulations for each configuration.  } \label{grapheff}
\begin{minipage}{0.47\textwidth}
    \includegraphics[width=1\textwidth]{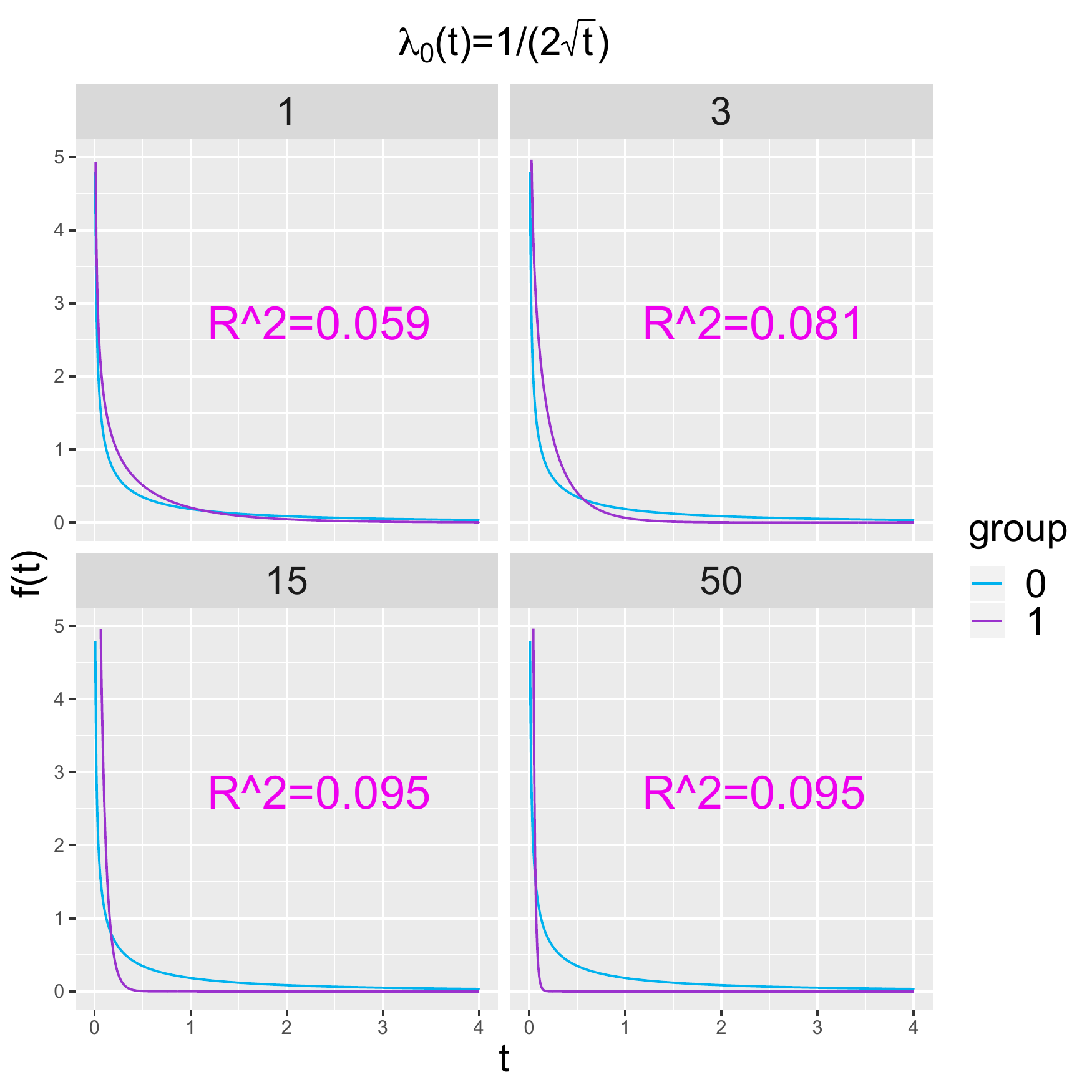}
    \end{minipage}
    \hspace{\fill} 
    \begin{minipage}{0.47\textwidth}
    \includegraphics[width=1\textwidth]{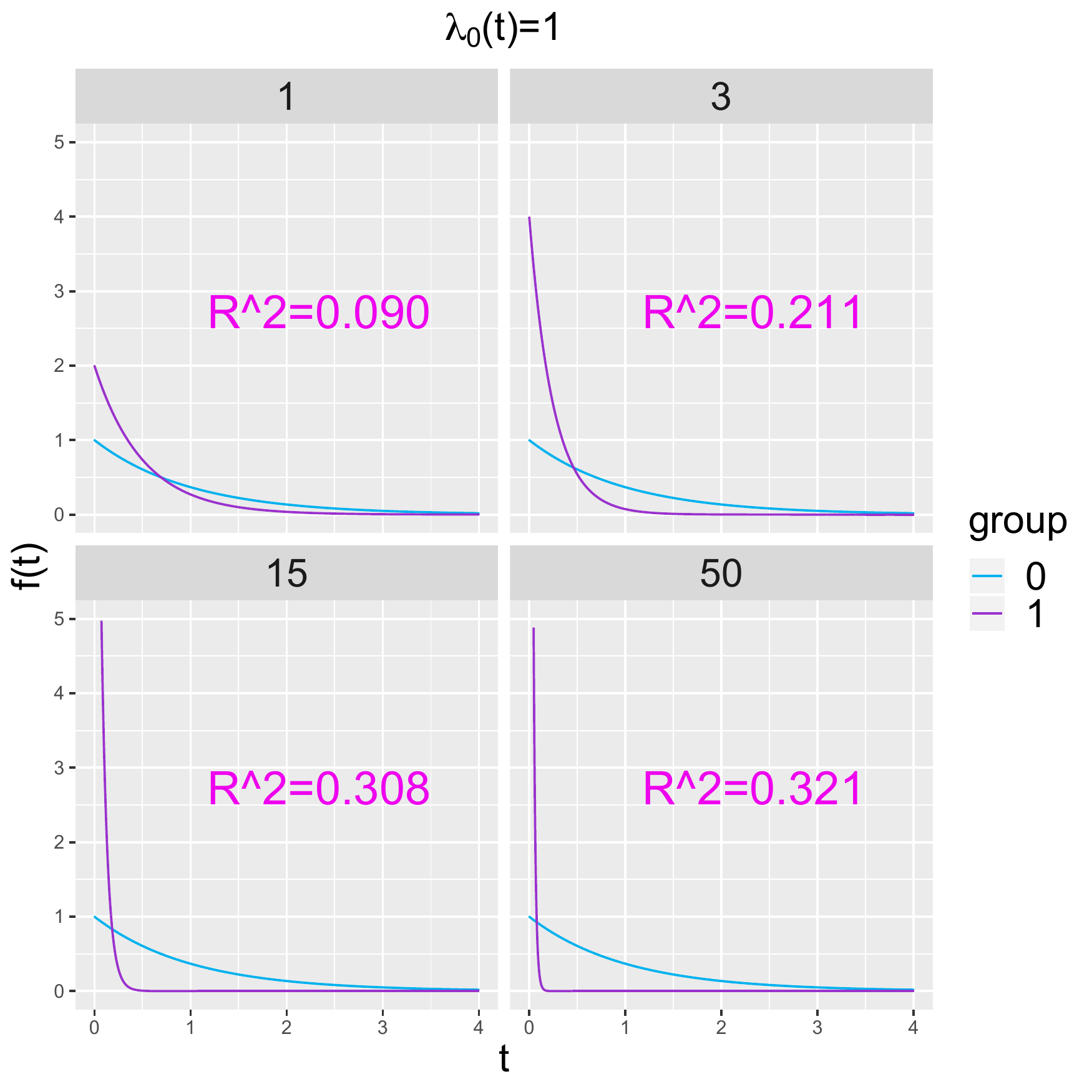}
    \end{minipage}
\vspace*{0.3cm} 
 \begin{minipage}{1\textwidth}
    \begin{center}
    \includegraphics[width=0.47\textwidth]{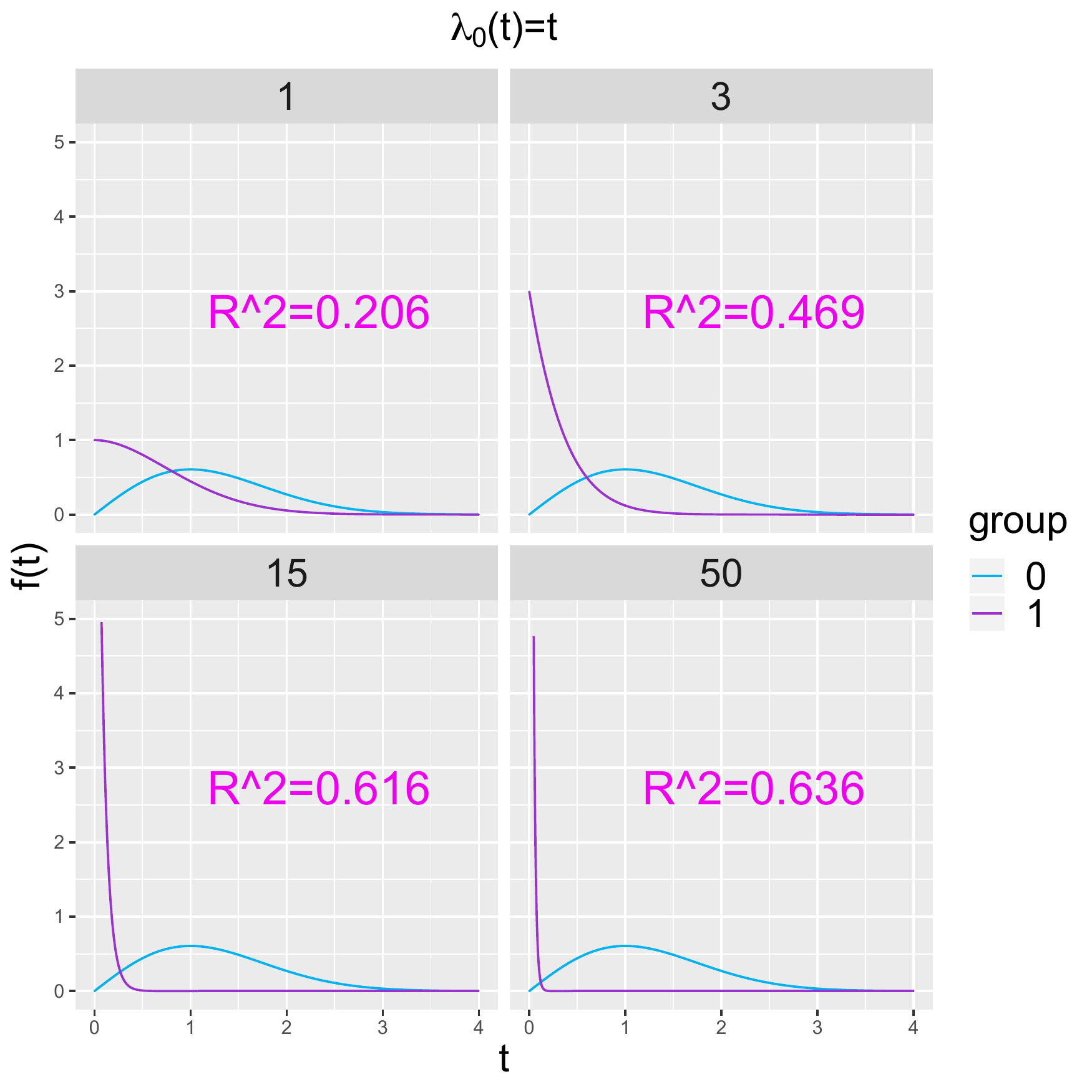}
    \end{center}
    \end{minipage}
\end{figure}

We consider here a binary $Z$ taking values 0,1 with equal probabilities. We  consider $\lambda_0(t)= 1$, $ t$ and $1/(2\sqrt{t})$.
In Figure \ref{grapheff} we plot the  density  of $T$ for each group, to show how the two groups differ in each scenario. 
The mean of $R^2_{\infty}$ over the 100 simulations are printed on each configuration.
From Figure \ref{grapheff} we see that the $R^2_{\infty}$ values tend to be larger when the two groups indexed by $Z=0$, 1 have different concentrations of failure times, i.e.~different shapes of the density functions, such as in the case of $\lambda_0(t)=t$.  
On the contrary, with $\lambda_0(t)= 1/(2\sqrt{t})$ the two density functions have very similar shapes, resulting much smaller  $R^2_{\infty}$ values. 
As noted earlier, the upper bound of $\Omega^2$ for the three cases are  0.091,  0.333 and 0.647, respectively.

\subsubsection*{Nested models}

Next we consider a limited set of simulations with data generated under $\lambda(t|Z)=\lambda_0(t) +Z_1+3Z_2+Z_3$, where the covariates $Z_1,Z_2$ and $Z_3$ were independently  drawn from Uniform $[0,\sqrt{3}]$ and the baseline hazard was in turn equal to $1,t$ and $1/(2\sqrt{t})$. 
We also consider an additional pure noise covariate $Z_4 \sim$  Uniform $[0,\sqrt{3}]$, not used in the data generating mechanism.
We consider the following models listed in Table \ref{tab:mult}: three univariate models with each of $Z_1,Z_2$ and $Z_3$, respectively;  a model with only $Z_1$ and $Z_3$; a model with all the three $Z_1,Z_2,Z_3$; and a model with the three covariates plus the pure noise $Z_4$.
We see from Table \ref{tab:mult} that $R^2_{\infty}$ increases with the complexity of the models: $R^2_{\infty}$ with both $Z_1$ and $Z_3$ is larger than with $Z_1$ or $Z_3$ alone; meanwhile, since $Z_2$ has a strong effect as reflected in its regression coefficient, $R^2_{\infty}$ with $Z_2$ alone is larger than with both $Z_1$ and $Z_3$. 
The measure is substantially larger with all three covariates $Z_1,Z_2$ and $Z_3$ than under any of the previous models.
With the noise variable $Z_4$ added to the model, $R^2_{\infty}$ increases very slightly from 0.122 to 0.124, for example. This also informs us how to use the $R^2$ type measures for model selection: if the addition of a variable only increases the $R^2$ very slightly, it is perhaps not worth the cost of an extra degree of freedom. This is consistent with the concept of adjusted $R^2$, which explicit adjusts for the number of degrees of freedom. We further discuss this in the applications later. 

\subsection{Comparison with the measure under the Cox Model}\label{cox}

 As discussed earlier the semiparametric additive hazards model behaves somewhat differently from the semiparametric Cox model. 
 Here we compare $R^2_\tau$ as defined in  \eqref{r2} under the two models when both models are valid. 
We consider a binary $Z$ and  constant baseline hazard; this is a case where both the semiparametric additive hazards model \eqref{add_const} and the classic Cox model hold. 

Under the Cox model  $S(t\mid Z)=\{S_0(t)\}^{\exp(\beta Z)}$,  where the regression parameter is typically estimated using the partial likelihood, and the baseline survival function via  the Breslow's estimate of the cumulative baseline hazard. 
We can then similarly estimate the explained variation as defined in \eqref{omega} or \eqref{omega_tau}, using a similar approach as described in Section 3.  We denote this as $R^2_{cox}$. Both $R^2_{cox}$ and $R^2_\tau$ thus defined  should be consistent for the same $\Omega^2_\tau$. 
In Table \ref{coxtable} 
 we again simulated with  $\lambda(t|Z)=1+\beta Z$ for a binary $Z$, $\beta =1,3,15$ and 50, with no censoring or $30\%$ censoring . 
As expected, the values of $R^2_{cox}$ and $R^2_\tau$ are indeed very close to each other.

\subsection{Explained variation of $Z$ given $T$}\label{ZgivenT}

\citet{oq:xu:12} advocated for considering the explained variation of $Z$ given $T$ under the Cox regression model. 
One main advantage of this approach is that the resulting measure tend not to be bounded strictly less than one. In addition, considering  $Z$ given $T$ is also consistent with the sequential conditioning and counting process notation often used in survival analysis. 
Following \citet{PNAS} and \citet{oq:xu:12}, 
 we  consider in particular the covariate residual (also called Schoenfeld residual under the Cox model) based approach. 

In order to obtain the residuals of $Z$, we need to estimate the conditional  distribution of $Z$ given $T$. 
A theorem from  \citet{ZgivenT, xu:oq} can be readily adapted to provide a consistent estimate of this conditional  distribution under model \eqref{add_const}:
\begin{thm}
Under model \eqref{add_const} and independent censoring, assuming that $\lambda_0(t)$ is known (or otherwise  consistently estimated), 
the conditional distribution of Z given T is consistently estimated by
\be
    \hat{P}(Z\leq z \mid T=t)=\frac{\sum_{Z_j\leq z} Y_j(t)\left(\lambda_0(t)+\hat{\beta}^TZ_j\right)}{\sum_{l=1}^n Y_l(t)\left(\lambda_0(t)+\hat{\beta}^TZ_l\right)}.
\ee
\end{thm}
The proof of the above theorem is similar to that of Theorem 1 in  \citet{ZgivenT, xu:oq} but applied to model \eqref{add_const}.

In practice $\lambda_0(t)$ is unknown, and also not readily estimated by the typical software that fit the additive hazards model. Our investigation here is of exploratory nature, aimed to understand the behaviors of the explained variation of $T$ give $Z$ versus $Z$ given $T$.  
In simulations below we  use the true $\lambda_0(t)$. 
 Denote
\be
\hat{E}_\beta\left(Z\mid t\right)=\frac{\sum_{j=1}^nZ_jY_j(t) \left(\lambda_0(t)+ \beta Z_j \right)}{\sum_{l=1}^nY_l(t) \left(\lambda_0(t)+ \beta Z_l \right)}.
\ee
The residuals under the fitted model and under the `null' model where $\beta=0$ are, respectively:
\be
r_i(\hat{\beta})=Z_i-\hat{E}_{\hat{\beta}}\left(Z\mid X_i\right), \,\,\,
r_i(0)=Z_i-\hat{E}_{0}\left(Z\mid X_i\right),
\ee
where $ {E}_{0}\left(Z\mid X_i\right) $ is simply the empirical average of $Z$ in the risk set at time $X_i$.
Therefore for a scalar $Z$ we may define
$$
R^2_{Z\mid T}=1-\frac{\sum_{i=1}^nr^2_i(\hat{\beta})}{\sum_{i=1}^nr^2_i(0)}.
$$
The extension to multivariate $Z$ was described in \citet{oq:xu:12} and can be easily adopted here. 

We simulated under
$
\lambda(t)=1+\beta Z 
$,
with a binary $Z$   and equal probabilities of 0, 1.
In Table \ref{resZT} we  see that unlike  $R^2_\tau$, the values of $R^2_{Z\mid T}$  approach one with 
increasing $|\beta|$.  
We further discuss the unknown $\lambda_0(t)$  in the last section.

\section{Applications}\label{seer-sec}


\subsection{Leukimia: FREIREICH DATA}

\begin{figure}[H]
\caption{Difference between the cumulative hazard functions of the two groups for the Freireich data. } \label{fre}
 \begin{center}
    \includegraphics[width=0.47\textwidth]{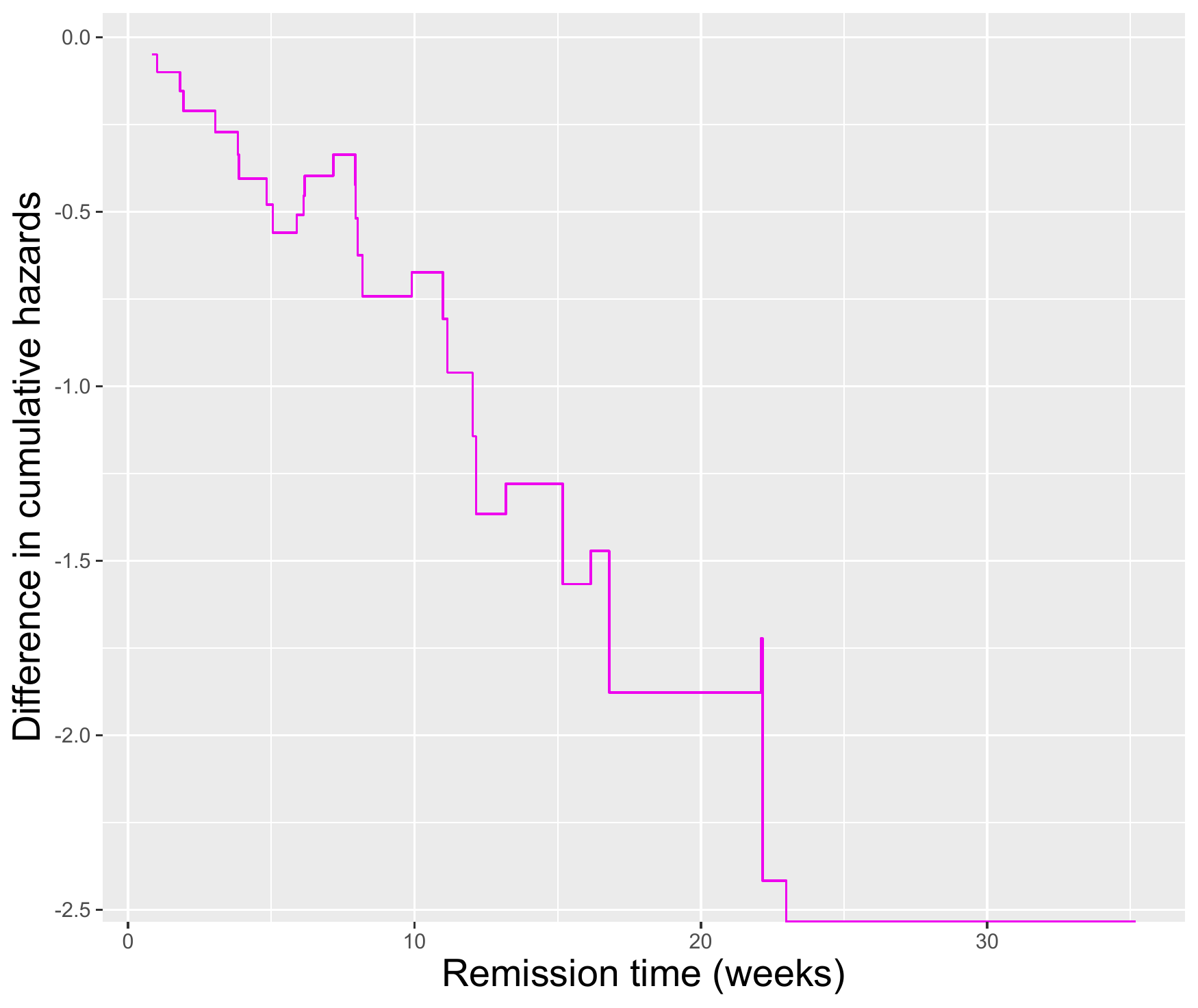}
\end{center}
\end{figure}

We first apply the measure of explained variation to the \citet{Leuk} data, 
which consist of  the remission times of 42 Leukimia patients in a randomized clinical trial treated with the drug 6-mercaptopurine (6-MP) versus placebo.  The data set has been well-known in the survival analysis literature, and was in the first table of   \citet{cox:oakes}. 
As a diagnostic plot in Figure \ref{fre} we show the difference of the cumulative hazard functions between the two treatment groups; 
under the semiparametric additive hazards model \eqref{add_const} this difference should be  linear in time. 
From the figure we see that except for random noise due to limited sample size the difference  shows a very nice linear trend, indicating that the semiparametric model \eqref{add_const} fits the data reasonably well.
We note that in the R package `timereg' that we used to fit the semiparametric additive hazards model, no diagnostic tools appear to be provided for checking this model. 

We calculated $R^2 = 0.201$, indicating, as is known, good separation between the two groups' survival times. Typically if a single predictor, in particular a binary one, turns out to have an $R^2$ of around 20\% say, it is considered to be a strong predictor.
Previously the explained variation of $Z$ given $T$ under the Cox regression model had been calculated to be around 0.40  (ranging from 0.38 to 0.42 depending on the  measure used) \cite{oq:xu:12}. 
The Freireich data appears to be a data set that fits both the proportional hazards model and the additive hazards model reasonably well. 
Based on the simulation results, when the data fits both models,  the explained variation of $T$ given $Z$ would be very close under the two models. 
The discrepancy between the $R^2$ values seen above are most likely attributable to the difference between 
 the explained variation of $Z$ given $T$ and that of $T$ given $Z$, as also illustrated in the simulations. In this case they otherwise reflect somewhat comparable strengths of association in our opinion.

\subsection{Prostate cancer: SEER-MEDICARE DATA}

We study the time to death of 29,657  prostate cancer patients with localized non-metastatic disease identified from the linked Surveillance, Epidemiology, and End Results (SEER) - Medicare database, diagnosed between 2004 and 2009. Following \citet{hou:par:hou:18} we consider the clinical and the demographical variables,  plus the binary insurance claims codes from Medicare. The latter captures medical diagnoses and procedures through 
Healthcare Common Procedure Coding System (HCPCS) codes, international classification of diseases (ICD)-9 diagnosis and procedure codes, etc. Each insurance claims code variable takes value one if that claim appeared within one year before diagnosis, and zero otherwise. 
Out of the 29,657  patients 3,543 died by the end of the follow-up which was December 2013 when the data were exported from the linked database.

The high dimensional data analysis of \citet{hou:par:hou:18} selected 143 variables to predict non-cancer mortality, and 9 variables to predict cancer mortality, in the context of these two competing risks. The same sets of variables were used in  \citet{riviere2019claims} and a complete list can be found in Table 1 and 2 of their supplemental material.
For our analysis of explained variation, we combined these two sets of predictor for overall survival, which resulted in  146 variables: PSA, Gleason Score, age, race (black versus other),  marital status (married versus other) and registry (California versus other), 
 plus the claims codes. A table with the distributions of these variables can be found in the Supplemental Materials.

In Figure \ref{seer}  we plot the difference of the cumulative hazard functions between groups as we did for the Freireich data above, to check the additive hazards model assumption. These are illustrated for 
 six binary variables, the three demographical variables plus three claims codes that are not too sparse to plot.
The plots  indicate that the model seems to fit the data reasonably well. 

We consider three models here. 
We first fit  the data to the semiparametric additive hazards model with only the cancer-related clinical variables PSA and Gleason Score. 
 We then add the four demographical variables. 
 Finally we added the set of  claim codes. 
The model fits are provided in the tables of the Supplement Materials. 
Table \ref{seer} summarizes the $R^2$ values obtained under these three models.  
In the first column of the table we see that the cancer-related clinical variables alone do not explain much  (under 1\%) variation in overall survival. This can at least be partially understood since only 734  out of the 3,543 total deaths in this data set were due to cancer.  
Demographical variables, on the other hand, do explain a substantial amount of variation in overall survival. 
This amount of explained variation is further  increased, by a non-trivial amount, after adding in the claims codes previously identified from the high-dimensional SEER-Medicare database. 

When high dimensional claims codes are used in the data analysis, there is often the concern of model over-fitting. In our case, with 3,543 death events and 146 total regressors, this may not be an issue. 
Nonetheless, 
we proceed to divide the data set randomly into two parts, a training set with 14,828 observations containing 1,803 deaths, and a test set with 14,829 observations containing 1740 deaths. 
We fit the additive hazards model to the training data set and obtain the estimates $\hat{\beta}$ and $\hat\Lambda_0(t)$. We use them to compute $\hat S (t|Z)$ on the test data set, and obtain
 an out-of-sample $R^2_{out}$. Such out-of-sample $R^2$ measures are often used in machine learning applications (eg.~deep learning) in order to reduce the risk of overfitting. We report the $R^2_{out}$ in Table \ref{seer}. 
 It is seen that, for this data, the $R^2_{out}$ values are in fact slightly higher than the $R^2$ computed on the full data set, or the $R^2_{train}$ computed on the training data set. Were there over-fitting, the $R^2_{out}$ values would have been substantially lower.
The discrepancy among the three quantities currently seen is mostly  due to variability in the estimation of the conditional survival function and consequently of the total and explained variances.
 For comparison purposes, we also provide in the Supplemental Materials the three model fits to the training data set. We can compare the estimated coefficients with those using the full data set, and observe that the estimates for the statistically signficant ones are stable across the training versus the full data set. 



At the suggestion of a reviewer, we compute the adjusted $R^2$, $R^2_{adj}=1- (1-R^2)(n-1)/(n-p-1)$, for the three models.  Here $n$ is the sample size, 
and $p$ is the number of the covariates included in the model. 
The $R^2_{adj}$ is computed on the full data set.
By definition $R^2_{adj} < R^2$, although no difference can be seen at three digits after the decimal point between the two measures for the first two models since $p$ is so small compared to $n$. For the third model that includes 146 variables, the difference of 0.3\% between the two does not appear to signify any over-fitting.

Finally we  note that 
the explained variation of $Z$ given $T$ under the Cox model, denoted $\rho^2$,  was calculated in \citet{riviere2019claims} for this data. They computed $\rho^2=0.71$ for cancer mortality and $\rho^2=0.60$ for non-cancer mortality under competing risks setting. 
As discussed before, the numerical values of explained variation of $T$ given $Z$ are not directly comparable to those  of $Z$ given $T$. Considering that the former has an upper bound less than one, it is perhaps also within reasons to  conclude that  our analysis under the additive hazards model agrees with that of \citet{riviere2019claims} about the contribution of the claims codes in explaining overall  mortality for this prostate cancer patient population.
This conclusion echoes the initial goal of the funded project that lead to the previous publications  \cite{hou:par:hou:18, riviere2019claims} to demonstrate that the high-dimensional insurance claims codes contain useful information about mortality in this  patient population.

\section{Discussion}

In this paper we have studied explained variation under the semiparametric additive hazards model for right-censored survival data. 
 The explained variation is shown to lie between zero and one, and to increase with the magnitude of the regression effect.
It has been known, and is shown again here, that the explained variation of survival time given covariates can have an upper bound  strictly less than one.
 Nonetheless, \citet{ash:shwa:99} argues convincingly that low $R^2$ values can be useful as a measure of model performance and prediction, and we have illustrated the same in our data analyses. Indeed in many of today's genome-wide association studies, polygenic risks scores are commonly assessed using $R^2$ measures, even though their values are typically very low (single digit of percentage points) for most diseases studied. 
 
 The semiparametric additive hazards model is different in several aspects from the historically more widely used semiparametric proportional hazards model.  The model and hence its inference is not rank invariant, which makes it less familiar to most users in the seimparametric survival analysis field. This phenomenon also carries over to the explained variation under the model, leading to its dependence on the baseline hazard function. Of course, the choice of a model should depend on how close it is to the true data  generating mechanism.  On the other hand, as mentioned earlier the semiparametric additive hazards model is known to be collapsible, 
 and this makes it more sensible to compare nested models which, as we have illustrated, is a common usage of $R^2$ type measures.


As reviewed in the Introduction, other approaches exist in the literature in order to develop $R^2$ type measures. In the Simulation section, we have considered a residual based approach, that relates to the explained variation of the covariates given the survival time. This was an approach advocated under the Cox proportional hazards model \cite{oq:xu:12}, as it does not encounter the problem of being bounded strictly less than one. Unfortunately, for the additive hazards model, it requires the knowledge or consistent estimation of the baseline hazard function $\lambda_0(t)$, which is not provided in the commonly used software such as the R package `timereg'.  Smoothing methods such as kernels may be applied to $\reallywidehat{\Lambda}_0(t)$, 
and can be potentially used here, but this is beyond the scope of this work.
A third  approach  is based on information gain, but as it turns out,  it also requires an estimate of $\lambda_0(t)$ under the additive hazards model.

The R package `timereg' also allows $\beta$ to vary with time, i.e.~$\beta(t)$ in place of $\beta$ in model \eqref{add_const}. It estimates the cumulative $B(t)=\int_0^t\beta(u)du$, together with $\Lambda_0(t) =\int_0^t\lambda_0(u)du$. It is possible to define an $R^2$ measure similar to what we have done in this paper; the computation is in fact simpler because the estimated conditional survival function $S(t|z)$ is a step function. 
To our best knowledge little experience exists in the literature to inform us when to use this more general nonparametric model versus the semiparametric model we have considered here. We  have noticed that the nonparametric model does not appear  suitable for the two data sets  in this paper. The Freireich data set appears to have too small a sample size to the fit the nonparametric model, in that the resulting estimates are extremely bumpy and have large variation. The SEER-Medicare data set, on the other hand, is so sparse in the design matrix (i.e.~many zero values for the claims codes), together with high percentage of censoring, that the resulting estimated $B(t)$ is practically constant zero. This is not difficult to see from the formula 
$
\hat{B}(t)=(\mathbf{Z}^\top \mathbf{Z})^{-1}\mathbf{Z}^\top \int_0^t d\mathbf{N}(u)$,
where $\mathbf{Z}=[Z_1,\ldots,Z_n]^\top$ and $\mathbf{N}(u)=[N_1(u),\ldots,N_n(u)]^\top$.

The $R^2$ measure of explained variation 
should not be confused with goodness-of-fit measures, although there are connections between these two concepts. 
\citet{chauvel2017survival} show that the population version of the explained variation under the proportional hazards model will increase with improvements of fit, and that the best model from a large class of models maximizes the explained variation. They consider this in a similar setting as $\beta(t)$ in the above; see also \citet{flan:oq:19}. 
However, due to issues in fitting $\beta(t)$ under the additive hazards model, we have not been able to observe a similar phenomenon. 
This would be worth future investigation once we are able to have a good estimate of $\beta(t)$, perhaps with smoothing techniques.

The $R^2$ measure developed in this work has been implemented  in the R package `R2Addhaz' and is publicly available on CRAN. 

\section*{Appendix. $\Omega^2$ increases with $\left| \beta \right|$: proof of a specific case}

Here we prove that $\Omega^2$ increases with $\left| \beta \right|$
when $Z$ is Bernoulii with $p= 0.5$ and  under the semiparametric hazards model \eqref{add_const}. We have:
\be
E\left\{\mbox{Var}\left(T\mid Z\right)\right\}&=& E\left\{E\left(T^2\mid Z\right)\right\}-\left[E\left\{E\left(T^2\mid Z\right)\right\}\right]^2 \nonumber\\
&=&\frac{1}{2}\left[2\int_0^\infty t \exp{\left\{-\Lambda_0(t)\right\}}dt - \left[\int_0^\infty \exp{\left\{ -\Lambda_0(t)\right\}}dt \right]^2\right]
\nonumber\\
&&+  \frac{1}{2}\left[2\int_0^\infty t \exp{\left\{ -\Lambda_0(t)-\beta t \right\}}dt - \left[\int_0^\infty \exp{\left\{ -\Lambda_0(t) -\beta t \right\}}dt \right]^2\right]
\ee
and
\be\label{vart}
\mbox{Var}(T)&=&E\left\{\mbox{Var}\left(T\mid Z\right)\right\}+\mbox{Var}\left\{E\left(T\mid Z\right)\right\} \nonumber\\
&=&E\left\{E\left(T^2\mid Z\right)\right\}-\left[E\left\{E\left(T\mid Z\right)\right\}\right]^2 \nonumber\\
&=&\frac{1}{2}\left[2\int_0^\infty t \exp{\left\{ -\Lambda_0(t)\right\}}dt +2\int_0^\infty t \exp{\left\{ -\Lambda_0(t)-\beta t \right\}}dt \right]
\nonumber\\
&&-\left[\frac{1}{2} \int_0^\infty \exp{\left\{ -\Lambda_0(t) \right\}}dt +\frac{1}{2}\int_0^\infty \exp{\left\{ -\Lambda_0(t) -\beta t \right\}}dt \right]^2.
\ee
If we take the derivative with respect to $|\beta|$ of these quantities we get:
\be\label{vartz}
\frac{\partial E\left\{\mbox{Var}\left(T\mid Z\right)\right\}}{\partial \left| \beta \right|}&=& 
-\mbox{sign}(\beta) \int_0^\infty t^2 \exp{\left\{ -\Lambda_0(t)-\beta t \right\}}dt+ \mbox{sign}(\beta)\int_0^\infty \exp{\left\{ -\Lambda_0(t) -\beta t \right\}}dt\int_0^\infty t\exp{\left\{ -\Lambda_0(t) -\beta t \right\}}dt\nonumber\\
&=& -\mbox{sign}(\beta)\frac{1}{3}E\left(T^3\mid Z=1\right)+\mbox{sign}(\beta)\frac{1}{2}E\left(T\mid Z=1\right)E\left(T^2\mid Z=1\right)
\ee
and
\be\label{vart}
\frac{\partial \mbox{Var}(T)}{\partial \left| \beta \right| }&=&-\mbox{sign}(\beta)\int_0^\infty t^2 \exp{\left\{ -\Lambda_0(t)-\beta t \right\}}dt  \nonumber\\
&&+\mbox{sign}(\beta)\left[\frac{1}{2} \int_0^\infty \exp{\left\{ -\Lambda_0(t) \right\}}dt +\frac{1}{2}\int_0^\infty \exp{\left\{ -\Lambda_0(t) -\beta t \right\}}dt \right]\int_0^\infty t\exp{\left\{ -\Lambda_0(t) -\beta t \right\}}dt 
\nonumber\\
&=& -\mbox{sign}(\beta) \frac{1}{3}E\left(T^3\mid Z=1\right)+\mbox{sign}(\beta)\frac{1}{2}\left[\frac{1}{2}E\left(T\mid Z=1\right)+\frac{1}{2}E\left(T\mid Z=0\right)\right]E\left(T^2\mid Z=1\right).
\ee
By equation \eqref{vartz} and \eqref{vart}, and after some algebra:
\be
\frac{\partial \Omega^2_\infty}{\partial \left| \beta \right|}&=&
-\mbox{sign}(\beta)\frac{E\left\{\mbox{Var}\left(T\mid Z\right)\right\}\left\{\frac{1}{4}E\left(T\mid Z=1\right)E\left(T^2\mid Z=1\right)-\frac{1}{4}E\left(T\mid Z=0\right)E\left(T^2\mid Z=1\right)\right\}}{\left(\mbox{Var}(T)\right)^2}
\nonumber\\
&&-\mbox{sign}(\beta)\frac{\mbox{Var}\left\{E\left(T\mid Z\right)\right\}\left\{-\frac{1}{3}E\left(T^3\mid Z=1\right)+\frac{1}{2}E\left(T\mid Z=1\right)E\left(T^2\mid Z=1\right)\right\}}{\left\{\mbox{Var}(T)\right\}^2}.
\ee

If  now we consider the special case of $\lambda_0(t)=1$, for which $\lambda(t)>0$ if and only if $\beta>-1$,  we have:
\be
\frac{\partial \Omega^2_\infty}{\partial \left| \beta \right|}&=&\frac{\mbox{sign}(\beta)}{\left\{\mbox{Var}(T)\right\}^2}\left[\frac{1}{4}E\left\{\mbox{Var}\left(T\mid Z\right)\right\}E\left(T^2\mid Z=1\right)\left(\frac{\beta}{1+\beta}\right)+\mbox{Var}\left\{E\left(T\mid Z\right)\right\}\left(\frac{1}{(1+\beta)^3}\right)\right]
\nonumber\\
&=&\frac{\left|\beta\right|\left(2+\beta\right)}{\left\{4\mbox{Var}(T)\right\}^2\left(1+\beta\right)^4}>0,
\ee
proving that the measure increases with $\left| \beta \right|$.

\bibliography{R2}

\newpage

\newpage

\begin{table}[H]
\caption{Simulation results for different values of $\beta$ and $\tau$ under the model $\lambda(t)=1+\beta Z$; in () are standard errors from simulation runs. }\label{fix}
\begin{tabular}{lrrrrrr}
\hline
\headrow
 \multicolumn{1}{c}{$\beta$ } & \multicolumn{1}{c}{$Z$} & \multicolumn{1}{c}{$\tau$} &\multicolumn{1}{c}{Censor} 
 & \multicolumn{1}{c}{$\reallywidehat{\beta}$ }   & \multicolumn{1}{c}{$R^2_\tau$} & \multicolumn{1}{c}{$\Omega^2_{\tau}$} \\
 \hiderowcolors
\rowcolor[gray]{1} &  &$\infty$ & $0\%$& $1.000$ $(0.117)$ & $0.072$ $(0.017)$ & $0.074$ \\
 & $U\left(0,\sqrt{3}\right)$ & $4.3$ & $14\%$ & $0.992$ $(0.128)$ & $ 0.067$ $(0.017)$ & $0.071$ \\
 &   & $1.3$ &$39\%$ & $ 0.994$ $(0.151)$ & $0.027$ $(0.013)$ & $0.026$\\
 \hiderowcolors
 \rowcolor[gray]{0.9} 
  \multirow{-2}{*}{1}
  &  &$\infty$ & $0\%$ & $1.000$ $(0.103) $ & $0.090$ $(0.014)$ & $0.090$ \\
 & Binary  & $4.3$ & $17\%$ & $ 1.001$ $(0.115)$ & $0.087$ $(0.017)$ &$0.090$ \\
  & & $1.3$ & $45\%$ &$  1.006$ $(0.140)$ & $ 0.029$ $(0.014)$ &$0.029$ \\
 \rowcolor[gray]{1}
  &  &$\infty$ &$0\%$& $2.996$ $(0.227)$ & $0.190$ $(0.027)$ & $0.191$ \\
  & $U\left(0,\sqrt{3}\right)$ & $4.3$ &$8\%$ &  $2.984$ $(0.238)$ & $ 0.186$ $(0.029)$ & $0.192$ \\
  &  & $1.3$ & $25\%$ & $ 2.997$ $(0.259)$ & $0.128$ $(0.027)$ &$0.129$ \\
  \hiderowcolors
 \rowcolor[gray]{0.9} \multirow{-2}{*}{3}
 & &$\infty$ & $0\%$&$3.020 \;(0.184)$ & $ 0.211$ $(0.018)$& $0.209 $\\
  & Binary & $4.3$ &$14\%$ &  $3.037$ $(0.231)$ & $ 0.234$ $(0.020)$ & $0.229$ \\
  &  & $1.3$ & $38\%$  & $ 2.962$ $(0.210)$ & $0.158$ $(0.033)$ &$0.166$\\
\rowcolor[gray]{1}
  &  &$\infty$ &$ 0\%$&$ 15.077$ $(0.765)$ & $ 0.368$ $(0.044)$  & $0.360$\\
  & $U\left(0,\sqrt{3}\right)$ & $4.3$ & $3\%$ & $15.175$ $(0.916)$ & $ 0.377$ $(0.047)$ &$0.367$ \\
 &   & $1.3$ & $10\%$  & $ 15.041$ $(0.925)$ & $0.356$ $(0.050)$ &$0.363$ \\
 \hiderowcolors
  \rowcolor[gray]{0.9}\multirow{-2}{*}{15} 
 & &$\infty$ & $0\%$ &$15.053$ $(0.765) $ & $0.308$ $(0.020)$ & $0.304$\\
  & Binary & $4.3$ &$12\%$  & $ 14.943$ $(0.669) $ & $ 0.353$ $(0.021)$ &$0.341$\\
 &  & $1.3$ & $30\%$  & $15.083$ $(0.781)$ & $0.431$ $(0.024)$ &$0.431$\\
 \rowcolor[gray]{1}
  &  &$\infty$ & $0\%$& $ 49.438$ $(3.272)$ & $0.438$ $(0.070)$ & $0.430$\\
   & $U\left(0,\sqrt{3}\right)$ & $4.3$ &$1\%$ &$49.878$ $(2.446)$ & $ 0.452$ $(0.069)$ &$0.440$\\
   & & $1.3$ &$4\%$  &$ 49.741$ $(2.695)$ & $ 0.456$ $(0.069)$ &$0.467$\\
 \hiderowcolors
  \rowcolor[gray]{0.9}\multirow{-2}{*}{50}
 &  & $\infty$ &$0\%$&$50.3373$ $(2.465) $ & $0.321$ $(0.020)$ & $0.324$\\
  & Binary & $4.3$ & $12\%$  & $49.761$ $(2.577) $ & $0.374$ $(0.018)$ &$0.364$\\
 & & $1.3$ & $29\%$ & $ 50.056$ $(2.479)$ & $0.486$ $(0.022)$ &$0.484$ \\
\hline  
\end{tabular}
\end{table}

\newpage

\newpage
\begin{table}[H]
\caption{$R^2$ values  for nested models; in () are standard errors from simulation runs. 
}\label{tab:mult}
\begin{tabular}{lcccc}
\hline 
\headrow
{Model} & \thead{$\lambda_0(t)=1$} & \thead{$\lambda_0(t)=t$} & \thead{$\lambda_0(t)=1/(2\sqrt{t})$}    \\ 
$Z_1$ & $0.012$ $(0.005)$ & $0.016$ $(0.008)$ & $0.008$ $(0.005)$   \\
$Z_2$ & $0.084$ $(0.019)$ & $0.122$ $(0.020)$ & $0.060$ $(0.015)$  \\
$Z_3$ & $0.013$ $(0.007)$ & $0.015$ $(0.006)$ & $0.008$ $(0.005)$\\
$Z_4$ & $0.001$ $(0.001)$ & $0.001$ $(0.001)$ & $0.001$ $(0.001)$\\
$Z_1+Z_3$& $0.025$ $(0.010)$ & $0.031$ $(0.011)$ & $0.017$ $(0.010)$\\
$Z_1+Z_2+Z_3$ & $0.122$ $(0.028)$ & $0.174$ $(0.031)$ & $0.087$ $(0.023)$ \\
$Z_1+Z_2+Z_3+Z_4$ & $0.124$ $(0.029)$ & $0.176$ $(0.031)$ & $0.089$ $(0.024)$ \\
\hline 
\end{tabular}
\end{table}

\begin{table}[H] 
\caption{
Comparison of explained variation under the semiparametric additive hazards model and the semiparametric Cox model, when both models are correct;  in () are standard errors from simulation runs.}\label{coxtable}
\begin{tabular}{lcrrr}
\hline 
\headrow
\thead{$\beta$} & {Censor} & \thead{$R^2_\tau$} & \thead{$R^2_{cox}$} \\
 \hiderowcolors
\rowcolor[gray]{1}
  & $0\%$ & $0.094$ $(0.015)$ & $0.094$ $(0.015)$\\
  \multirow{-2}{*}{1}
   & $30\%$  & $0.063$ $(0.018)$ & $0.063$ $(0.016)$\\
   \rowcolor[gray]{0.9}
  & $0\%$ & $ 0.208$ $(0.015)$ & $0.208$ $(0.015)$\\
  \multirow{-2}{*}{3}
   & $30\%$ & $ 0.207$ $(0.027)$ & $0.208$ $(0.026)$\\
  \rowcolor[gray]{1}
& $0\%$ &    $0.306$ $(0.022)$ & $0.306$ $(0.022)$\\
 \multirow{-2}{*}{15}
 & $30\%$ &    $0.433$ $(0.025)$ & $0.434$ $(0.025)$\\
 \rowcolor[gray]{0.9}
& $0\%$ &  $0.329$ $(0.022)$ & $0.330$ $(0.022)$\\
 \multirow{-2}{*}{50}
 & $30\%$ &  $0.491$ $(0.023)$ & $0.493$ $(0.023)$\\
\hline
\end{tabular}
\end{table}

\begin{table}[H]
\caption{Explained variation of $Z|T$ versus $T|Z$;  in () are standard errors from simulation runs.  }\label{resZT}
\begin{tabular}{lc|rrrrrr|}
\hiderowcolors
\hline
 \cellcolor[gray]{0.8}{$\beta\;\;\;\;\;\;$} & 1 & \cellcolor[gray]{0.9}{3} & 15 & \cellcolor[gray]{0.9}{50} & 100 & \cellcolor[gray]{0.9}{1000}\\
\cellcolor[gray]{0.8}$R^2_{Z \mid T}$  & $0.099$ $(0.020)$ & \cellcolor[gray]{0.9}$ 0.291$ $(0.024)$  & $ 0.668$ $(0.026)$  & \cellcolor[gray]{0.9}$ 0.851$ $(0.020)$ & $0.911$ $(0.017)$ & \cellcolor[gray]{0.9} $0.988$ $(0.006)$\\
\cellcolor[gray]{0.8}$R^2_{\infty}$  & $0.090$ $(0.016)$ & \cellcolor[gray]{0.9}$ 0.208$ $(0.017)$  & $ 0.308$ $(0.022)$  & \cellcolor[gray]{0.9}$ 0.328$ $(0.021)$ & $0.332$ $(0.021)$ & \cellcolor[gray]{0.9} $0.333$ $(0.020)$\\ \hline
\end{tabular}
\end{table}

\newpage

\begin{figure}[H]
\caption{Difference between the cumulative hazard functions of groups defined by some dichotomous variables for the SEER-MEDICARE data. } \label{seer}
\begin{minipage}{0.4\textwidth}
    \includegraphics[width=1\textwidth]{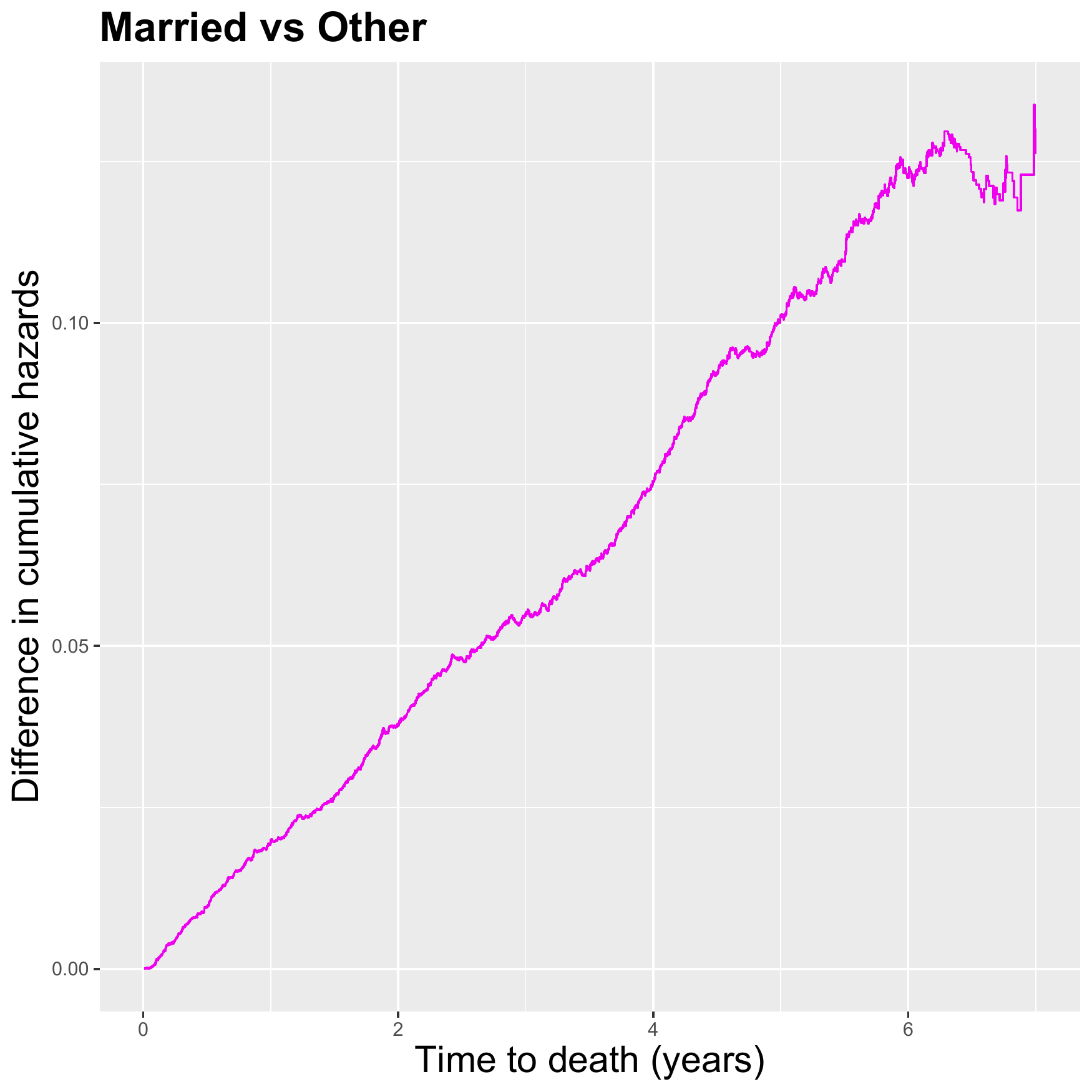}
    \end{minipage}
    \hspace{\fill} 
    \begin{minipage}{0.4\textwidth}
    \includegraphics[width=1\textwidth]{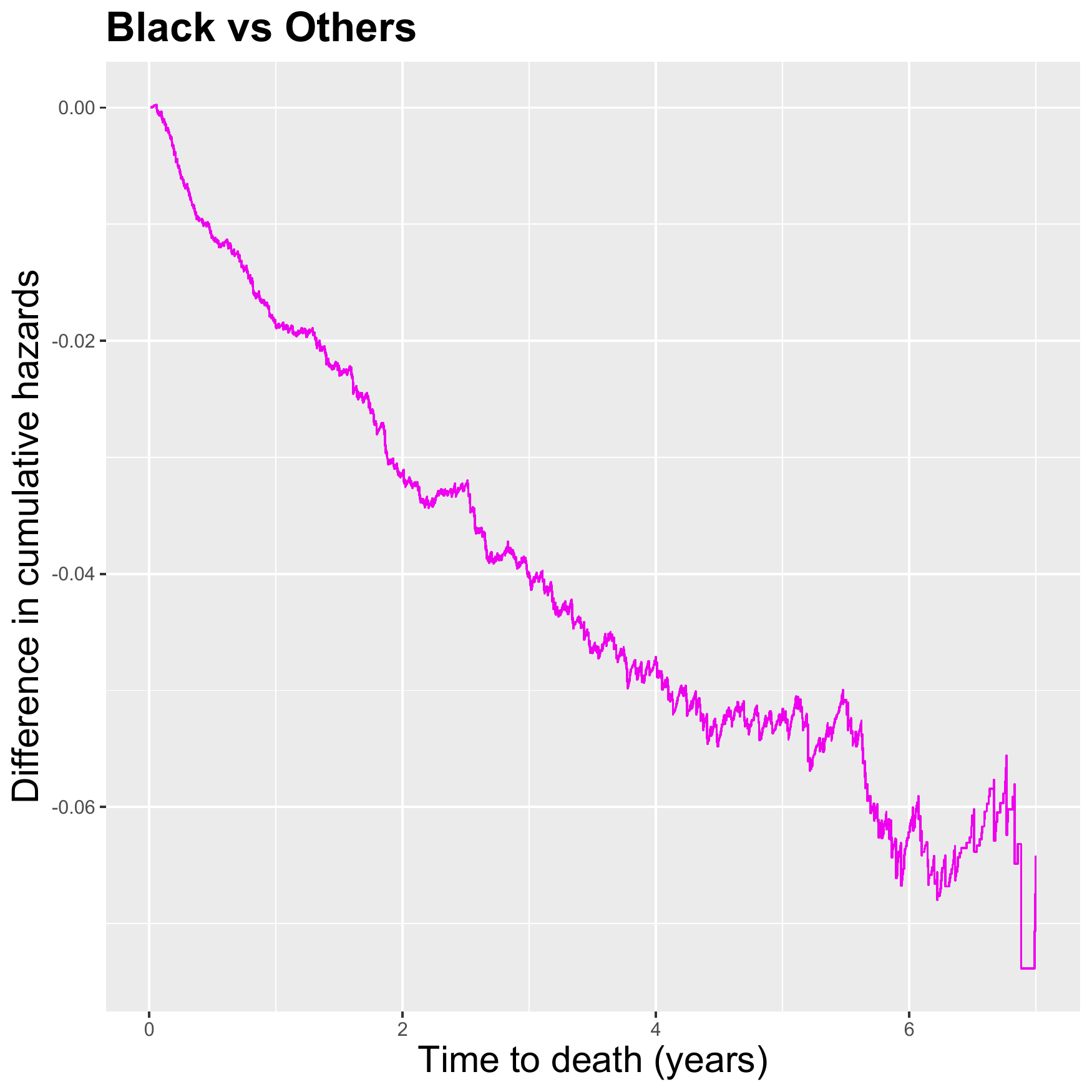}
    \end{minipage}
    \hspace{\fill} 
         \vspace*{0.3cm}
    \begin{minipage}{0.4\textwidth}
    \includegraphics[width=1\textwidth]{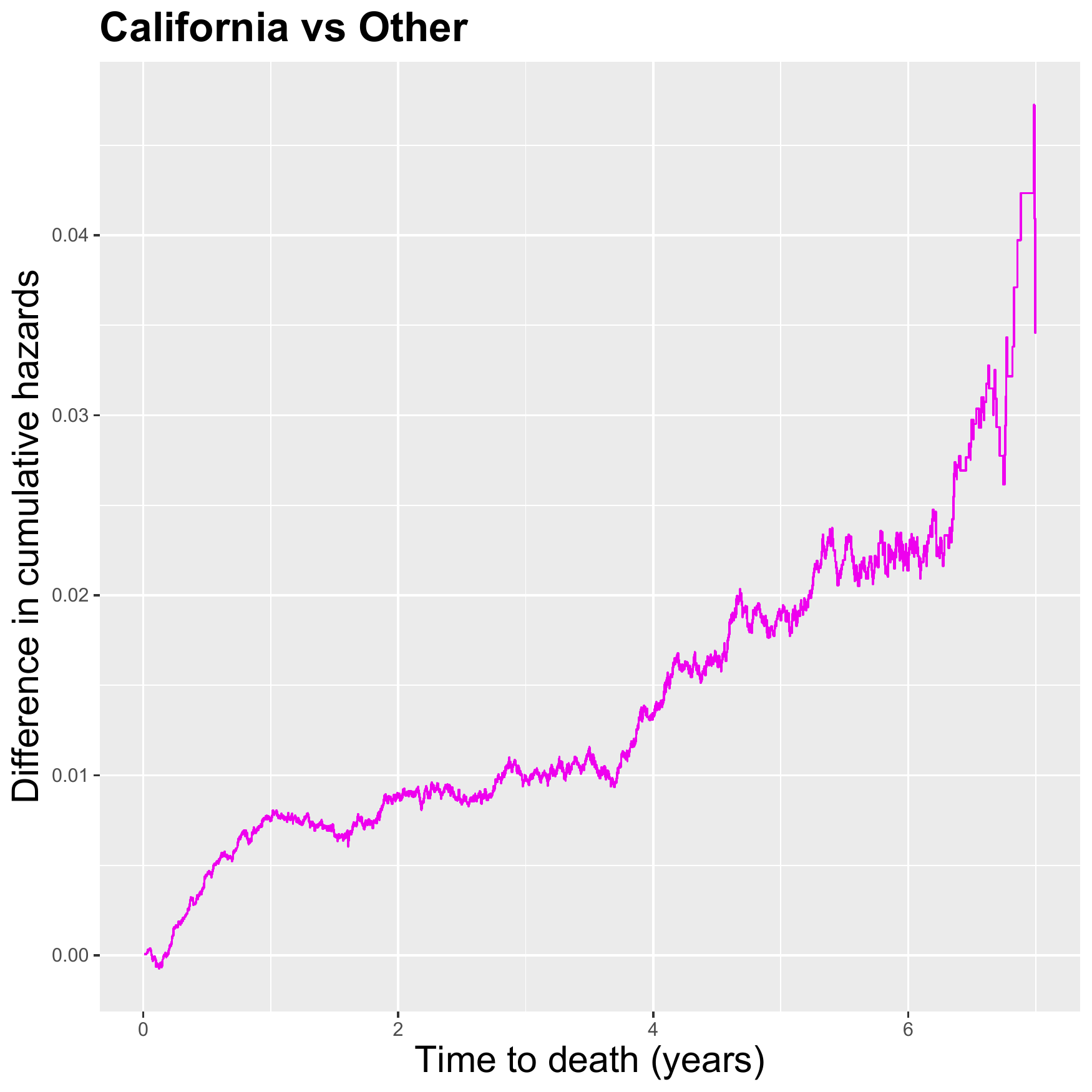}
    \end{minipage}
     \hspace{\fill}
    \begin{minipage}{0.4\textwidth}
    \includegraphics[width=1\textwidth]{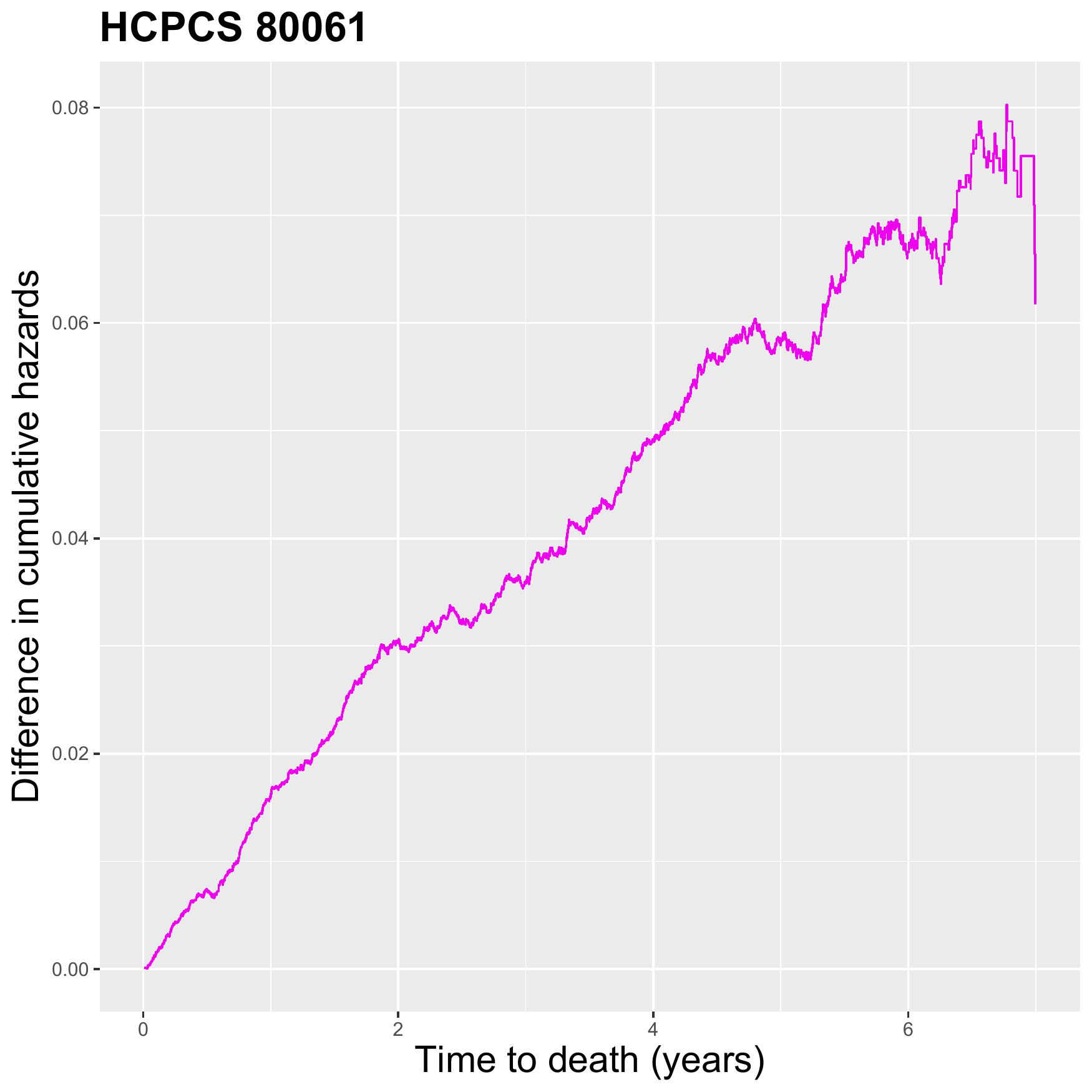}
    \end{minipage}
 \hspace{\fill} 
      \vspace*{0.3cm}
 \begin{minipage}{0.4\textwidth}
    \includegraphics[width=1\textwidth]{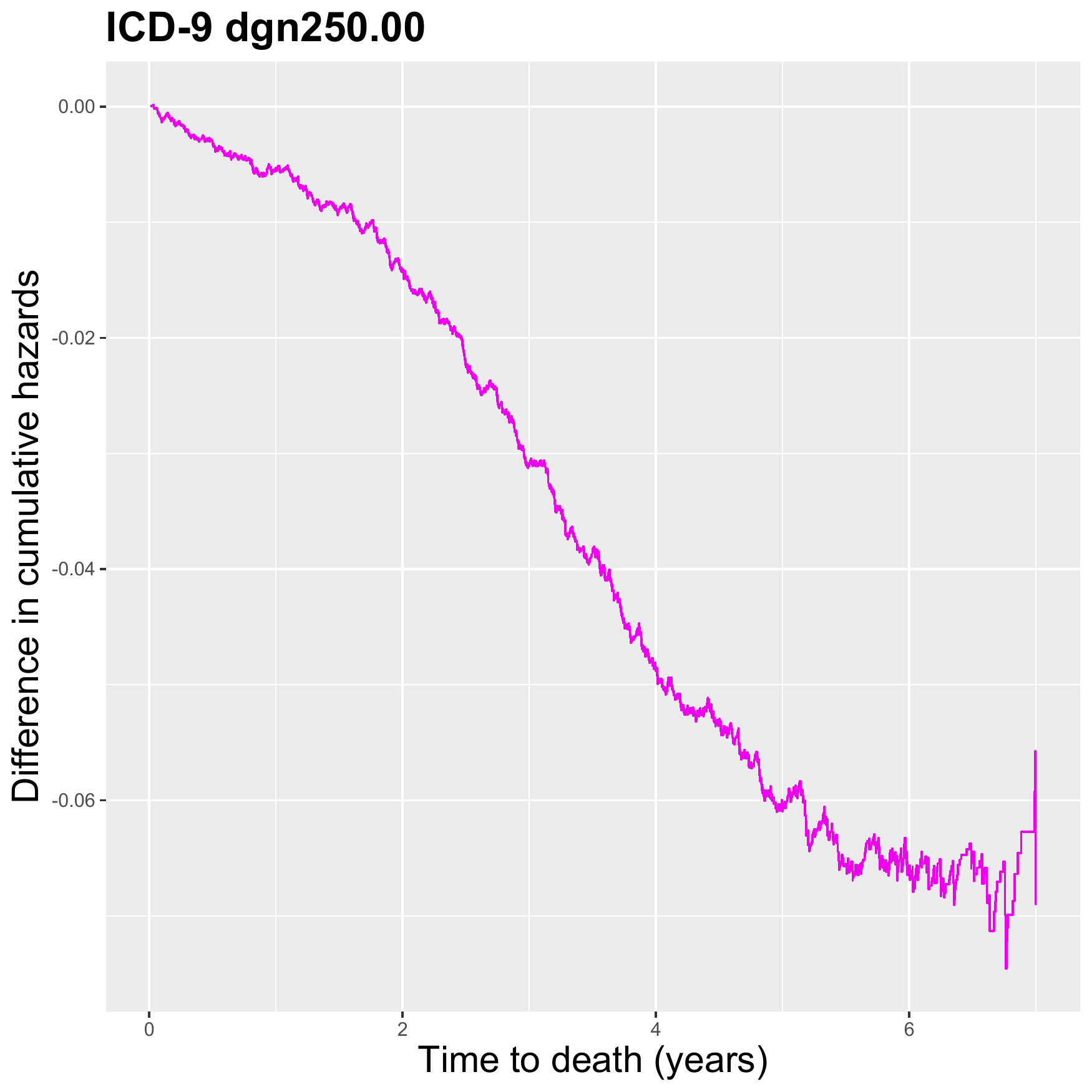}
    \end{minipage}
    \hspace{\fill}  
    \begin{minipage}{0.4\textwidth}
    \includegraphics[width=1\textwidth]{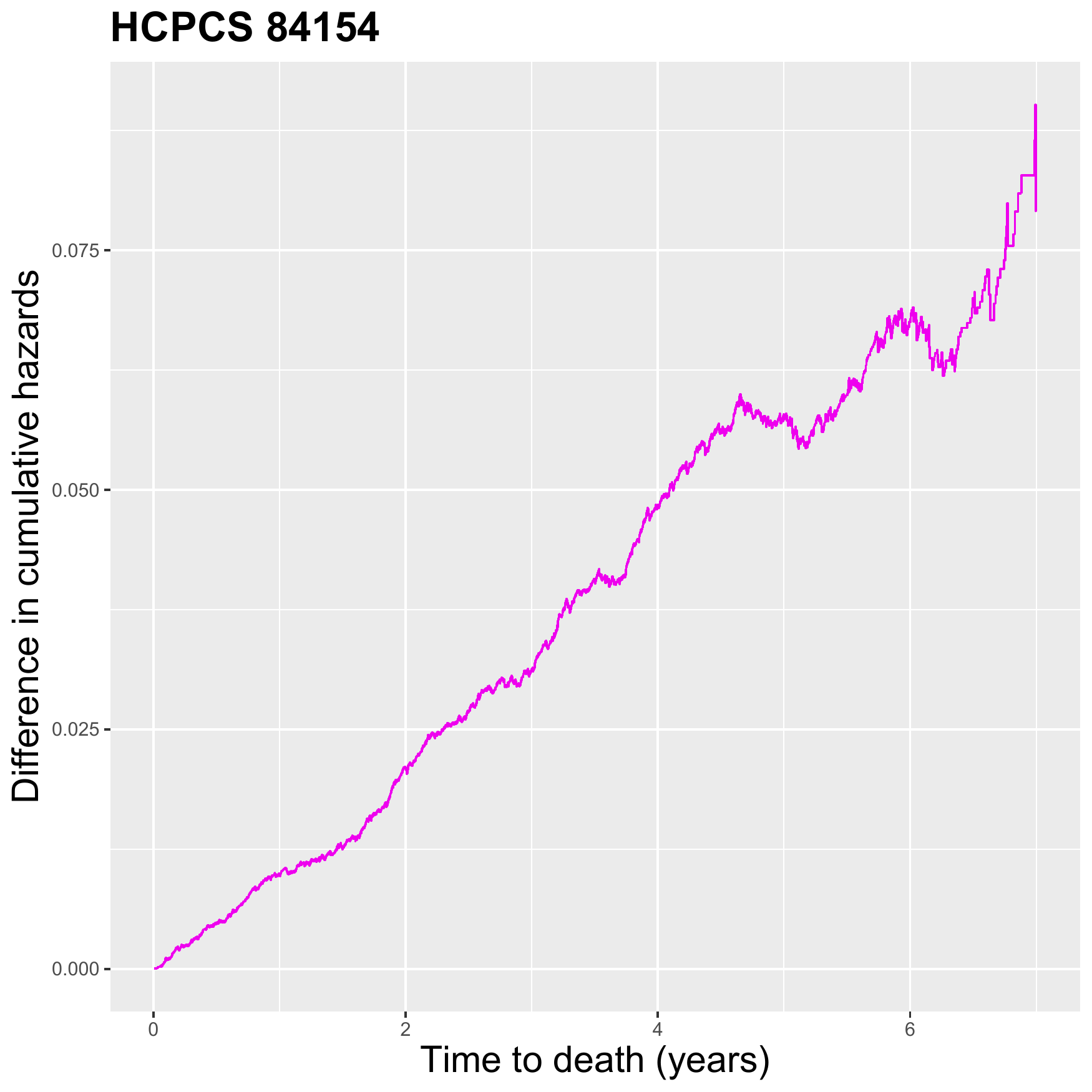}
    \end{minipage}
\end{figure}

\newpage

\newpage
\begin{table}[H]
\caption{$R^2$ values  for the SEER-Medicare data set. $R^2$ is computed on the full data set; $R^2_{adj}$ is the adjusted $R^2$ also computed on the full data set; $R^2_{out}$ is the out-of-sample $R^2$ computed on the test data set, with all parameters estimated from the training data set;  
and $R^2_{train}$ is computed only on the training data set. 
}\label{seer}
\label{mult}
\begin{tabular}{lccccc}
\hline 
\headrow
{Model} & \thead{$R^2$} & \thead{$R^2_{adj}$} & \thead{$R^2_{out}$}   & \thead{$R^2_{train}$}   \\ 
Clinical & $0.048$  & $0.048$ & $0.053$ & $0.051$\\
Clinical + Demo. & $0.270$  & $0.270$ & $0.271$  & $0.261$  \\
Clinical + Demo. + Claims & $0.373$ & $0.370$ & $0.388$ & $0.379$ \\
\hline 
\end{tabular}
\end{table}

\end{document}